# Linear and nonlinear analysis of Ion-Temperature-Gradient (ITG) Driven mode in the asymmetric Pair-Ion Magnetoplasma

Javaria Razzaq[1], Zahida Ehsan[2,3] and Arshad M. Mirza[1]

[1] *Theoretical Plasma Physics Group, Physics Department, Quaid-i-Azam University, Islamabad, 45320, Pakistan.*

[2] *SPAR and The Landau-Feynmann Laboratory for Theoretical Physics, CUI, Lahore Campus 54000, Pakistan.* and

[3] *National Centre for Physics(NCP), Shahdara Valley Road, Islamabad 45320, Pakistan.*[*]

## Abstract

We have investigated linear and nonlinear dynamics of ion-temperature-gradient driven drift mode for Maxwellian and non Maxwellian pair-ion plasma embedded in an inhomogeneous magnetic field having gradients in ion's temperature and number density. Linear dispersion relations are derived and analyzed analytically as well as numerically for different cases. It has been found that growth rate of instability increases with increasing $\eta$.By using the transport equations of Braginskii, model, a set of nonlinear equations are derived. In the nonlinear regime, soliton structures are found to exist. Our numerical analysis shows that amplitude of solitary waves increases by increasing ion to electron number density ratio. These solitary structures are also found to be sensitive to asyemmtries in pair plasma and non thermal kappa and Cairns distributed electrons. Our present work may contribute a good illustration of the observation of nonlinear solitary waves driven by the ITG mode in magnetically confined pair-ion plasmas and space pair-ion plasmas as the formation of localized structures along drift modes is one of the striking reasons for L-H transition in the region of improved confinements in magnetically confined devices like tokamaks.

[*]Electronic address: ehsan.zahida@gmail.com

## I. BACKGROUND

Lepton epoch, which in physical cosmology is believed to have happened one second after the big bang is an era where electron positron plasma created by the annihilation of hadrons and anti-hadrons were dominating the universe. Of course these pairs of plasma particles were in thermal equilibrium with electromagnetic radiations (photon gas).

Pair plasmas possess unique features of time-space parity because they can be mobile in the electromagnetic fields in the same way due to their equal mass and opposite charge of same strength. Therefore electron positrons can provide excellent opportunity to understand the matter and anti-matter research [1–4, 6**?** , 7].

Curiosity of the scientists to dig deep into the early universe and processes happening at stellar distances has been the first building block of laboratory astrophysics. In this regard, importance of positrons physics has been well recognized in high energy physics in particular to understand the properties of antimatter such charge, parity and time reversal (CPT) invariance, Bose-Einstein condensates and production of positrons in the laboratory is carried out for the better understanding of blackholes, gamma ray bursts like astrophysical environments.

Due to annihilation of electron positron, it is very difficult to maintain the ep-plasma for a longer period of time in the laboratory. However, in pair ion plasma we do not have such problems of annihilation. To mimic pair plasma in a controlled enviornmet, in Japanese scientists Hatekayama and Oohara succesfully craeted lighter pair ion plasma (hydrogen ion) plasma and high density Fullerene ion ($C_+^{60}$ and $C_-^{60}$ ) plasma. In parallel, some results obtain by the experiments need support from theoreis, in this regard a kinetic theory taking into account the boundary effects [8] has been developed.However, still a theory is required which can support experimental observations and basic fluid theory.

The first observations where temperature of both electrons ($T_e$) and positrons ($T_p$) was reported by Chen et al., where author reported $T_p \lesssim$ (0.5 $T_e$ the effective electrons temperature) where effective temperature of positrons: $T_p$ was found to be ˜ $2.8\pm0.3$MeV[9]. On the other hand at Fermi National Accelerator Laboratory, Batavia, experiments from Tevatron collider indicated an unbalanced proportion of matter to antimatter where matter appears to dominate the antimatter beyond the limit of 1% predicted by the Standard Model. These observation clearly indicated asymmetries in both temperature and mass of the pair plasmas.

Plasma at electron positron collider created by using slightly different Lorentz factor beams is also asymmetric. Interactions between the charged particles or some naturally occuring nonlinear process can also lead to mass asymmetries whereas it can also assumed as initial condition[10].

Temperature and mass asymmetries in plasma have been responsible for the onset of many instabilities (such as Weibel like instabilities in case of temperature asymmetry) and other nonlinear phenomena however some contradictory studies also exist in literature.. For a pair ion plasma case, Verheest et al. [11] demonstrated that a strict symmetry destroys the stationary nonlinear structures of acoustic nature and showed such nonlinear structures can exist when there is a thermodynamic asymmetry between both constituents. Whereas in context of propagation of electromagnetic waves in the asymmetric pair plasma, Mahajan et al., assuming $T_- = T_+$ derived the nonlinear Schrödinger equation (NLSE) in which the nonlinear term vanishes when the temperatures of the electrons and positrons are equal [2] however this was found to be not consistent with the earlier results by Shukla et al. and Tajima et al.,[12–14].

Recently Ehsan et al., [7] reported a new acoustic-like mode which exists only for asymmetric or non isothermal pair plasma model when $T_- \ll T_+$. The real and imaginary parts are given as below:

$$\omega_r = \frac{k\left(\frac{T_+}{m_-}\right)^{1/2}}{\left(1+k^2\lambda_{D+}^2\right)^{1/2}} \tag{1}$$

and

$$\omega_i = -\sqrt{\frac{\pi}{8}}\frac{k\left(\frac{T_+}{m_-}\right)^{1/2}\left(\frac{m_+}{m_-}\right)^{1/2}}{\left(1+k^2\lambda_{D+}^2\right)^2}\left[1+\left(\frac{T_+}{T_-}\right)^{3/2}\left(\frac{m_-}{m_+}\right)^{1/2}\exp\left(-\frac{1}{2}\frac{\left(\frac{T_+}{T_-}\right)}{\left(1+k^2\lambda_{D+}^2\right)}\right)\right] \tag{2}$$

where $\omega_\pm = \left(\frac{4\pi n_\pm e^2}{m_\pm}\right)^{1/2}$. This is a heavy ion a low frequency branch of longitudinal oscillations in the pair-ion plasma system. For $|\omega_i| \ll \omega_r$, the temperature ratio of the plasma components give:

$$\left(\frac{T_+}{T_-}\right)^{3/2}\exp\left(-\frac{(T_+/T_-)}{2(1+k^2\lambda_{D+}^2)}\right) \ll 1 \tag{3}$$

On the other hand, enormous literature on various type of drift wave instabilities to solve plasma confinement problem exists for such instabilities are considered to be responsible for anomalous energy transport and enhanced particle diffusion. For instance in tokamak and space plasmas much attention has been given towards ion (electron) temperature gradient (I(E)TG) for their vital role in the ions (electrons) anomalous thermal transport. ITG mode arises due to inhomogenieties in temperature and density owing to the linear coupling of ion sound and drift wave. Needless to mention initially ITG mode instability was investigated for uniform density plasma[15], later on density inhomogeniety and magnetic shear were also incorporated[16].

In this manuscript we plan to see how a new longitudinal mode identified for the pure pair plasma couples with the drift waves and excites nonlinear stationary structures. This will be intriguing to see how inhomoegneity in temperature, density and then asymmetries in mass (temperature) of the pair particles impact ITG instability. Since in temperature and mass asymmetries will enhance the complexities in the system and the associated differential equations therefore their significant role on the ITG instability growth rate will be expected.

Here we shall study different situations, first when there are asymmetries in both temperature and mass of the positive and negative ions [7]. Then we shall see how presence of electrons population along with asymmetries in positive and ions as indicated by Saleem[17] and Verheest can affect the instability and subsequent nonlinear structures. 3) We will examine the above cases for Kappa and Cairnes distributed plasmas[18, 19].

To the best of the authors' knowledge, this has not been studied earlier and the results of the present investigation are useful to study nonlinear structures driven by ITG modes.

This paper is organized in the following manner: In Sec. II, the basic set of equations for the ITG mode in the asymmetric pair plasma have been presented. Sections III & IV deal with the derivation of linear dispersion relation for Maxwellian asymmetric pure pair plasma and with the presence of electrons, respectively. Whereas nonlinear analysis for both the cases has been carried out in the subsections.. Sections IV provides linear and nonlinear analysis of ITG mode for the non-Maxwellian pair ion plasma. while in Sec. V, numerical results are presented and summarized with a brief discussion of main results.

## II. BASIC SET OF EQUATIONS

We consider a system of collisionless plasma which contains $s$ species embedded in an inhomogeneous external magnetic field $\mathbf{B}_o = B_o(x)\hat{z}$ (where $\hat{z}$ is the unit vector along the $z - axis$). The gradients in temperature ($\nabla T_{so}(x)$) and density ($\nabla n_{so}(x)$), are assumed to be along the x-axis.

The ion momentum equation for the $s$th species, can be written as

$$m_s n_s \left(\frac{\partial}{\partial t} + \mathbf{v}_s \cdot \nabla\right) \mathbf{v}_s = -Z_s e n_s (\mathbf{E} + \frac{1}{c}\mathbf{v}_s \times \mathbf{B}_0) - \nabla (n_s T_s) \tag{4}$$

where $\mathbf{E} = -\nabla\phi$ ($\phi$ is the electrostatic wave potential), the subscripts $s$ stands for species. $m_s$, $Z_s$, $T_s$, and $n_{s0}$ are the mass, the charge state, temperature and number density of the $sth$ species, respectively. The perpendicular fluid velocity for low-frequency $\partial_t \ll \omega_{cs,}$ electrostatic waves (where $\omega_{cs} = Z_s e B_o / m_s c$ is the standard gyrofrequency) can be written as $\mathbf{v}_{s\perp} \approx \mathbf{v}_{EB} + \mathbf{v}_{Ds}$,where, $\mathbf{v}_{EB} = \frac{c}{B_o}\hat{z} \times \nabla\phi$, $\mathbf{v}_{Ds} = \frac{c}{eB_0 n_s}\hat{z} \times \nabla (n_s T_s)$ are the well known $\mathbf{E} \times \mathbf{B}$, and diamagnetic drift velocities, respectively.

The continuity equation is given as

$$\frac{\partial n_s}{\partial t} + \nabla \cdot (n_s \mathbf{v}_s) = 0, \tag{5}$$

the parallel component of ion fluid velocity can be expressed as

$$m_s n_s \left(\frac{\partial}{\partial t} + \mathbf{v}_s \cdot \nabla\right) v_{sz} = -Z_s n_s e \frac{\partial}{\partial z}\phi - \frac{\partial}{\partial z} (n_s T_s) \tag{6}$$

whereas the energy balance equation is given as:

$$\frac{3}{2} n_s \left(\frac{\partial}{\partial t} + \mathbf{v}_s \cdot \nabla\right) T_s + n_s T_s (\nabla \cdot \mathbf{v}_s) = -\nabla \cdot \mathbf{q}_s \tag{7}$$

where $\mathbf{q}_s$ is the Righi-Leduce ion heat flux which is defined as $(5cn_s T_s / 2Z_s e B_0)\hat{z} \times \nabla T_s$.

To close our system of equations, we shall use the Poisson's equation given by

$$\nabla^2 \phi = -4\pi Z_s e n_{s1}$$

Now to derive nonlinear set of mode coupling equations for the ITG driven drift mode in this case we write

$$n_{\pm 1} = n_{\pm} - n_{\pm o}(x)$$

$$T_{\pm 1} = T_{\pm} - T_{\pm o}(x)$$

where $n_{-1}(x)$ ($\ll n_{-o}$) and $T_{-1}(x)$ ($\ll T_{-o}$) are perturbed (equilibrium) number density and temperature, respectively. To derive nonlinear set of mode coupling equations for the ion-temperature-gradient driven drift mode for the pair-ion electron magnetoplasma, we may let $n_{s1} = n_s - n_{so}(x)$ and $T_{s1} = T_s - T_{so}(x)$, where $n_{s1}(x)\,(\ll n_{so})$ and $T_{s1}(x)\,(\ll T_{so})$ are perturbed (equilibrium) number density and temperature, respectively.

As the singly charged negative, positive ions and electrons are assumed to be in thermal equilibrium, so we can described their densities by the following expressions:

$$n_s(\phi) = n_{s0}\exp\left(\frac{e\phi}{k_B T_s}\right), \tag{8}$$

$$n_{i_+}(\phi) = n_{i_+0}\exp\left(-\frac{e\phi}{k_B T_{i_+}}\right) \tag{9}$$

for the case when electrons, singly charged negative and positive ions are taken as non-thermal, we can expressed the number densities for Kappa distributed electrons and singly charged negative and positive ions by the following expressions

$$n_s(\phi) = n_{s0}\left\{1 - \left(\kappa - \frac{3}{2}\right)^{-1}\frac{e\phi}{k_B T_s}\right\}^{-\kappa+1/2}, \tag{10}$$

$$n_{i_+}(\phi) = n_{i_+0}\left\{1 + \left(\kappa - \frac{3}{2}\right)^{-1}\frac{e\phi}{k_B T_{i_+}}\right\}^{-\kappa+1/2} \tag{11}$$

and for the case of Cairns distributed electrons and singly charged negative and positive ions densities are given by [18, 19]

$$n_s(\phi) = n_{s0}\left\{1 + \Gamma\left(-\frac{e\phi}{k_B T_s}\right) + \Gamma\left(-\frac{e\phi}{k_B T_s}\right)^2\right\}\exp\left(\frac{e\phi}{k_B T_s}\right), \tag{12}$$

$$n_{i_+}(\phi) = n_{i_+0}\left\{1 + \Gamma\left(\frac{e\phi}{k_B T_{i_+}}\right) + \Gamma\left(\frac{e\phi}{k_B T_{i_+}}\right)^2\right\}\exp\left(-\frac{e\phi}{k_B T_{i_+}}\right) \tag{13}$$

where the subscript $s$ equals $e$ for electrons and $i_-$ for negatively charged ions. Here, $\Gamma = 4\alpha/(1+3\alpha)$ comprise of Cairns parameter $\alpha$ that determines the population of nonthermal particles. It is important to mention here that if we consider $\kappa \to \infty$ and $\Gamma \to 0$ in Eqs. $(10-13)$, respectively, then we can directly replicate the Maxwell-Boltzmann density distribution for electrons, positively and negatively charged ions.

## III. CASE 1:(NEGATIVE AND POSITIVE IONS)

First we consider a case when there is a temperature and mass asymmetry in a two component plasma consisting of positive and negative ions, for this it is assumed that $T_\pm($ $m_\pm$) is the temperature (mass) of lighter and heavier ions. In this case, we assume $T_+ \gg T_-$ and $m_- \gg m_+$ so dynamics of heavier negative ions (given by Eq. (5-7)) will be taken into account while lighter ions are assumed to behave as Boltmanian.

Using drift-approximation, which is valid for low-frequency waves ($\omega \ll \omega_{c-,}$), the negative ion continuity, parallel component of ion momentum and energy balance equation for the $-$ ion species (Eq. (5-7)) can be expressed in the normalized form as:

$$(\zeta_t + \mathbf{v}_{B-}\cdot\nabla)\,N_- + Z_-\tau_-\,(\mathbf{v}_{B-} - \mathbf{v}_{n-})\cdot\nabla\Phi + \mathbf{v}_{B-}\cdot\nabla T_- + \frac{\partial}{\partial z}v_{-z} = 0 \tag{14}$$

$$(\zeta_t + v_{-z}\frac{\partial}{\partial z})v_{-z} = -v_{t-}^2\tau_-\frac{\partial}{\partial z}\left(Z_-\Phi + (N_- + T_-)\right) \tag{15}$$

$$(\zeta_t + \frac{5}{3}\mathbf{v}_{B-}\cdot\nabla)T_- - \frac{2}{3}\zeta_t N_- - Z_-\tau_-(\eta_- - \frac{2}{3})\mathbf{v}_{n-}\cdot\nabla\Phi = 0 \tag{16}$$

where $\zeta_t = \partial_t + (\mathbf{v}_{EB} + \mathbf{v}_{Ds})\cdot\nabla + v_{sz}\partial_z$, $\Phi = e\phi/T_+$, $v_{t-}^2 = T_-/m_-$ and $\tau_- = T_{+o}/T_-$.

### A. Linear Dispersion Relation

To obtain a linear dispersion relation, we shall drop all nonlinear terms and assume that the perturbed quantities $N_{-1}$,$T_{-1}$, $v_{-z}$ and $\Phi$ are proportional to $\exp[i\,(k_y y + k_z z - \omega t)]$, where $\omega$ and $(k_y, k_z)$ are perturbation frequency and wave vectors in the $y$ and $z$directions, then we obtain from Eq. (14)–(16)

$$(\omega - \omega_{B-})\,N_- - Z_-\tau_-\,(\omega_{n-} + \omega_{B-})\,\Phi - \omega_{B-}T_- - k_z v_{-z} = 0 \tag{17}$$

and

$$\omega v_{-z} = v_{t-}^2\tau_- k_z\left(Z_-\Phi + (N_- + T_-)\right) \tag{18}$$

and

$$\left(\omega - \frac{5}{3}\omega_{B-}\right)T_- - \frac{2}{3}\omega N_- + Z_-\tau_-\left(\eta_- - \frac{2}{3}\right)\omega_{n-}\Phi = 0 \tag{19}$$

where $\omega_{B-} = \mathbf{k}\cdot\mathbf{v}_{B-}$, and $\omega_{n-} = \mathbf{k}\cdot\mathbf{v}_{n-}$. Eliminating $T_-$ and $v_{-z}$ from Eq. (18) and Eq. (19) and substituting in Eq. (17), we get

$$\omega\left(\omega - \frac{5}{3}\omega_{B-}\right)\left(\omega - \omega_{B-} - \frac{k_z^2 v_{t-}^2}{Z_s\tau_-\omega}\right) - \omega\left(\omega - \frac{5}{3}\omega_{B-}\right)Z_-\tau_-\,(\omega_{n-} + \omega_{B-}) +$$
$$\omega Z_-\tau_-\omega_{n-}\omega_{B-}\left(\frac{2}{3} - \eta_-\right) - \left(\omega - \frac{5}{3}\omega_{B-}\right)\frac{k_z^2 v_{t-}^2}{Z_-} + k_z^2 v_{t-}^2\left(\eta_- - \frac{2}{3}\right)\omega_{n-} \tag{20}$$

Eq. (20) is the third order dispersion relation of ion-temperature-gradient drift mode for a plasma containing only positive and negative ions.

### B. Nonlinear Solution

A possible stationary solution of nonlinear set of Eqs. (14)-(16) can be obtained by introducing a new frame $\xi = y + \alpha z - ut$, where $\alpha$ and $u$ are constant and we assume that all the normalized perturbed quantities like $\Phi$, $N_-$, $v_{-z}$ and $T_-$ are only functions of $x$ and $\xi$ variables. In this new frame, the ion energy balance Eq. (16) yields the following result,

$$T_- = \frac{2}{3}N_- - \tau_- \left(\eta_- - \frac{2}{3}\right) \frac{\tilde{V}_{n-}}{u}\Phi \tag{21}$$

where $\tilde{V}_{n-} = \mathbf{v}_{n-} \cdot \hat{y}$.

Now, we transform the parallel component of ion fluid velocity Eq. (15), using the stationary frame coordinate transformation we get

$$\left(-u\frac{\partial}{\partial\xi} + \frac{c}{B_o}\hat{z} \times \nabla\phi \cdot \nabla\right) v_{-z} + v_{-z}\alpha\frac{\partial}{\partial\xi}v_{-z} = -v_{t-}^2\tau_-\alpha\frac{\partial}{\partial\xi}\left(Z_-\Phi + \Phi + T_-\right)$$

substituting the value of $T_-$

$$\begin{aligned} &-u\frac{\partial}{\partial\xi}v_{-z} + \frac{c}{B_o}\left(\frac{\partial\phi}{\partial x}\frac{\partial}{\partial\xi}v_{-z} - \frac{\partial\phi}{\partial\xi}\frac{\partial}{\partial x}v_{-z}\right) + v_{-z}\alpha\frac{\partial}{\partial\xi}v_{-z} = \\ &-v_{t-}^2\tau_-\alpha\frac{\partial}{\partial\xi}(Z_-\Phi + \Phi) - v_{t-}^2\tau_-\alpha\frac{\partial}{\partial\xi}\frac{2}{3}N_- - \tau_-\left(\eta_- - \frac{2}{3}\right)\frac{\tilde{V}_{n-}}{u}\Phi \end{aligned}$$

after simplifying we obtain the following result

$$\begin{aligned} v_{-z} = {} & \frac{\alpha v_{t-}^2}{u}\left[\left(1 - \left(\eta_- - \frac{2}{3}\right)\frac{\tilde{V}_{n-}}{u}\right)\Phi + \frac{5}{3\tau_-}N_-\right] \\ & + \frac{\alpha^3 v_{t-}^4}{2u^3}\left[\left(1 - \left(\eta_- - \frac{2}{3}\right)\frac{\tilde{V}_{n-}}{u}\right)\Phi + \frac{5}{3\tau_-}N_-\right]^2 \end{aligned} \tag{22}$$

similarly equation of continuity takes the following form

$$\frac{\partial}{\partial_\xi}\Phi - \frac{\tilde{V}_{B-}}{u}\frac{\partial}{\partial_\xi}\Phi - \frac{\tau_-}{u}\tilde{V}_{B-}\frac{\partial}{\partial_\xi}\Phi - \frac{\tau_-}{u}\tilde{V}_{n-}\frac{\partial}{\partial_\xi}\Phi + \alpha\frac{\partial}{\partial_\xi}v_{-z} = 0 \tag{23}$$

substituting the value of $v_{-z}$ and solving this system of nonlinear equation we obtain

$$a_1\frac{\partial\Phi}{\partial\xi} + a_2\Phi\frac{\partial\Phi}{\partial\xi} = 0$$

which represents shock like drift wave. Here

$$a_1 = 1 - \frac{\tilde{V}_{B-}}{u}(1+\tau_-) - \frac{\tau_-}{u}\tilde{V}_{n-} + \frac{\alpha v_{t-}^2}{u}\left\{1 - \left(\eta_- - \frac{2}{3}\right)\frac{\tilde{V}_{n-}}{u} + \frac{5}{3\tau_-}\right\}$$

$$a_2 = \frac{\alpha^3 v_{t-}^4}{2u^3}\left\{1 - \left(\eta_- - \frac{2}{3}\right)^2\left(\frac{\tilde{V}_{n-}}{u}\right)^2 + \left(\frac{5}{3\tau_-}\right)^2 - 2\left(\eta_- - \frac{2}{3}\right)\frac{\tilde{V}_{n-}}{u} + \frac{20}{3\tau_-}\right\}$$

## IV. CASE 2 (PRESENCE OF ELECTRONS)

Observed greater value than the expected one of the ion acoustic mode frequency in the experiments performed by Ohara[8] lead the scientists believe presence of electrons which are not fully filtered out from the chamber[17]. Therefore we discuss a case when electrons are also present in the pair plasma. For this $T_e > T_-, T_+$; $m_+ < m_-; T_- < T_+$

Using drift-approximation, which is valid for low-frequency waves ($\omega \ll \omega_{cs,}$), the pair-ion continuity equation, the parallel component of ion momentum equation and energy balance equation for the $s$th-species of ions can be expressed in the normalized form as:

$$(\zeta_t + \mathbf{v}_{Bs}\cdot\nabla)N_s + Z_s\tau_s(\mathbf{v}_{Bs} - \mathbf{v}_{ns})\cdot\nabla\Phi + \mathbf{v}_{Bs}\cdot\nabla T_s + \frac{\partial}{\partial_z}v_{sz} = 0 \tag{24}$$

$$(\zeta_t + v_{sz}\frac{\partial}{\partial_z})v_{sz} = -c_{ss}^2\frac{\partial}{\partial_z}\left(Z_s\Phi + \tau_s^{-1}(N_s + T_s)\right) \tag{25}$$

$$(\zeta_t + \frac{5}{3}\mathbf{v}_{Bs}\cdot\nabla)T_s - \frac{2}{3}\zeta_t N_s - Z_s\tau_s(\eta_s - \frac{2}{3})\mathbf{v}_{ns}\cdot\nabla\Phi = 0 \tag{26}$$

$\mathbf{v}_{Bs} = -\frac{cT_{so}}{Z_s e B_o}\hat{z}\times\nabla\ln B_o$, $\mathbf{v}_{ns} = -\frac{cT_{so}}{Z_s e B_o}\hat{z}\times\nabla\ln n_{so}$, are the standard $\nabla B_o$ and ion $\nabla n_{so}$ drifts, respectively. Here, $\eta_s = L_{ns}/L_{Ts}$ with $L_{ns} = 1/\left|\partial_x \ln n_{so}\right|$ and $L_{Ts} = 1/\left|\partial_x \ln T_{so}\right|$ denote the equilibrium ion density and temperature gradient scale lengths, respectively. The normalized parameters used in the above set of Eqs. (24) to (26) are defined as: $N_s = n_{s1}/n_{s0}$, $T_s = T_{s1}/T_{s0}$, $\Phi = e\phi/T_e$, represents the perturbed number density, temperature, electrostatic potential, respectively. Furthermore, $\tau_s = T_e/T_{s0}$ and $c_{ss} = \sqrt{T_e/m_s}$, is the ion-acoustic speed.

### A. Linear analysis

To obtain a linear dispersion relation, we shall drop all nonlinear terms and assume that the perturbed quantities $N_{s1}$,$T_{s1}$, $v_{sz}$ and $\Phi$ are proportional to $\exp[i(k_y y + k_z z - \omega t)]$, where $\omega$ and $(k_y, k_z)$ are perturbation frequency and wave vectors in the $y$ and $z$directions. Assuming plane-wave solution, Eqs. (24)-(26) can be re-written as

$$\left(\omega - \omega_{Bs}\right) N_s - Z_s \tau_s \left(\omega_{ns} + \omega_{Bs}\right) \Phi - \omega_{Bs} T_s - k_z v_{sz} = 0 \tag{27}$$

$$\omega v_{sz} = c_{ss}^2 k_z \left(Z_s \Phi + \tau_s^{-1} \left(N_s + T_s\right)\right) \tag{28}$$

$$(\omega - \frac{5}{3}\omega_{Bs})T_s - \frac{2}{3}\omega N_s + Z_s \tau_s (\eta_s - \frac{2}{3})\omega_{ns}\Phi = 0 \tag{29}$$

Here, we define $\omega_{Bs} = \mathbf{k} \cdot \mathbf{v}_{Bs}$, and $\omega_{ns} = \mathbf{k} \cdot \mathbf{v}_{ns}$. Eliminating $T_s$ and $v_s$ from Eq. (28)-Eq. (29), we get

$$\left(\omega - \omega_{Bs} - \frac{k_z^2 c_{ss}^2}{Z_s \tau_s \omega}\right) N_s = \left[Z_s \tau_s \left(\omega_{ns} + \omega_{Bs}\right) - \frac{Z_s \tau_s \omega_{ns} \omega_{Bs}}{(\omega - \frac{5}{3}\omega_{Bs})} \left(\frac{2}{3} - \eta_s\right) \right.$$
$$\left. + \frac{k_z^2 c_{ss}^2}{Z_s \omega} \left(1 - \frac{Z_s (\eta_s - \frac{2}{3})\omega_{ns}}{(\omega - \frac{5}{3}\omega_{Bs})}\right)\right] \Phi \tag{30}$$

$$\left(1 + k^2 \lambda_{De}^2\right) \Phi = Z_+ N_+ (\frac{n_{+0}}{n_{e0}}) - Z_- N_- (\frac{n_{-0}}{n_{e0}}) \tag{31}$$

where $\lambda_{De} = \sqrt{T_e / 4\pi n_{eo} e^2}$ is the electron Debye length. Eq. (30) relates the ion number density of the $s$th species with the normalized electrostatic potential. To close the system of equations, we may use the normalized form of Poisson's equation expressed by (31). Solving Eqs. (30) and (31) for pair-ion plasma, we obtain the following dispersion relation under the assumption that the perturbation wavelength is much larger than the electron Debye shielding length such that $k\lambda_{De} \ll 1$, and obtain the following relation,

$$\left[\omega - \frac{5}{3}\omega_{B+}\right] [\omega - \frac{5}{3}\omega_{B-}][Z_+ \tau_+ \omega^2 - Z_+ \tau_+ \omega_{B+} \omega - k_z^2 c_{s+}^2][Z_- \tau_- \omega^2 - Z_- \tau_- \omega_{B-} \omega - k_z^2 c_{s-}^2]$$
$$= Z_+ \alpha_o \left[\left\{\begin{array}{c} \tau_+^2 Z_+^2 \left(\omega_{n+} + \omega_{B+}\right) \left(\omega - \frac{5}{3}\omega_{B+}\right) \omega - \tau_+^2 Z_+^2 \omega_{n+} \omega_{B+} \left(\eta_+ - \frac{2}{3}\right) \omega + \\ k_z^2 c_{s+}^2 \tau_+ \left(\omega - \frac{5}{3}\omega_{B+} - Z_+ \omega_{n+} \left(\eta_+ - \frac{2}{3}\right)\right) \\ \left(Z_- \tau_- \omega^2 - Z_- \tau_- \omega_{B-} \omega - k_z^2 c_{s-}^2\right)(\omega - \frac{5}{3}\omega_{B-}) \end{array}\right\}\right]$$
$$+ Z_- \beta \left[\left\{\begin{array}{c} \tau_-^2 Z_-^2 \left(\omega_{n-} + \omega_{B-}\right) \left(\omega - \frac{5}{3}\omega_{B-}\right) \omega - \tau_-^2 Z_-^2 \omega_{n-} \omega_{B-} \left(\eta_- - \frac{2}{3}\right) \omega + \\ k_z^2 c_{s-}^2 \tau_- \left(\omega - \frac{5}{3}\omega_{B-} - Z_- \omega_{n-} \left(\eta_- - \frac{2}{3}\right)\right) \\ \left(Z_+ \tau_+ \omega^2 - Z_+ \tau_+ \omega_{B+} \omega - k_z^2 c_{s+}^2\right)(\omega - \frac{5}{3}\omega_{B+}) \end{array}\right\}\right] \tag{32}$$

where $Z_+n_{+o}(Z_-n_{-o})$ satisfies the quasineutrality condition of the plasma such that we can write $Z_+n_{+o} - Z_-n_{-o} = n_{eo}$, for pair-ion electron system. Here $\alpha_o = n_{+o}/n_{eo}$ and $\beta = [(1 - Z_+\alpha_o)/Z_-]$. Eq. (32) represents a sixth order dispersion relation and its analytical solution is extremely difficult to obtain. We shall therefore discuss numerical results in Sec. IV. We shall first present here, some interesting limiting cases in the following sub-section.

### 1. *Case-1:*

First, we consider a uniform density and magnetic field case, for which we can take $\omega_{n+} = \omega_{n-} = 0$; $\omega_{B+} = \omega_{B-} = 0$,and $\eta_+ = \eta_- = 0$. If we further assume that the phase velocity of wave lies in the rage of ion-acoustic and ion thermal speed such that: $v_{ts}^2 < \omega^2/k^2 < c_{ss}^2$ , we obtain the following dispersion relation:

$$\omega^2 = k_z^2 \left[\alpha_0 c_{s+}^2 + \beta c_{s-}^2\right] \tag{33}$$

where $\alpha_0$ and $\beta$ are defined earlier. The above dispersion relation represents a modified coupled ion-acoustic mode with variable mass, charge state and number density of positive and negative ions.

### 2. *Case-2:*

Secondly, we consider a nonunif

orm pair-ion plasma having equilibrium density and temperature gradients such that $\omega_{ns}, \neq 0, \eta_s \neq 0$. To investigate coupled ion-temperature-gradient- driven drift mode with ion-acoustic mode for pair-ion plasma, we assume the phase velocity of the wave in the range $c_{ss}^2 > \omega^2/k^2 > v_{ts}^2$ and keeping magnetic field uniform for simplification, we obtain

$$\begin{aligned}\omega^3 = k_z^2 \left[\alpha_o c_{s+}^2 \omega + \beta c_{s-}^2 \omega\right] + Z_+\alpha_o \left(\omega^2 \tau_+ Z_+ \omega_{n+} + c_{s+}^2 k_z^2 \omega_{n+} \left(\eta_+ - \frac{2}{3}\right)\right) \\ + Z_-\beta \left(\omega^2 \tau_- Z_- \omega_{n-} + c_{s-}^2 k_z^2 \omega_{n-} \left(\eta_- - \frac{2}{3}\right)\right)\end{aligned} \tag{34}$$

which is third order dispersion relation, essentially representing a coupled ion-acoustic type mode with ion-temperature-gradient driven drift mode.

## B. Nonlinear analysis

In the following section, we shall present a possible stationary solution of nonlinear set of Eqs. (24)-(26) by introducing a new frame $\xi = y + \alpha z - ut$, where $\alpha$ and $u$ are constant and we assume that all the normalized perturbed quantities like $\Phi$, $N_s$, $v_{sz}$ and $T_s$ are only functions of $x$ and $\xi$ variables. In this new frame, the ion energy balance Eq. (26) yields the following result,

$$T_s = \frac{2}{3}N_s - \tau_s\left(\eta_s - \frac{2}{3}\right)\frac{u_{ns}}{u}\Phi \tag{35}$$

where $\tilde{V}_{ns} = \mathbf{v}_{ns}\cdot\hat{y}$.

Now, we transform the parallel component of ion fluid velocity Eq. (25), using the stationary frame coordinate transformation and obtain the following result

$$v_{sz} = \frac{\alpha c_{ss}^2}{u}\left[\left(1 - \left(\eta_s - \frac{2}{3}\right)\frac{\tilde{V}_{ns}}{u}\right)\Phi + \frac{5}{3\tau_s}N_s\right] + \frac{\alpha^3 c_{ss}^4}{2u^3}\left[\left(1 - \left(\eta_s - \frac{2}{3}\right)\frac{\tilde{V}_{ns}}{u}\right)\Phi + \frac{5}{3\tau_s}N_s\right]^2 \tag{36}$$

Similarly the ion continuity equation can be written as,

$$\left(\partial_t + \frac{c}{B_o}\hat{z}\times\nabla\phi\cdot\nabla\right)n_{s1} - \frac{c}{B_o}\hat{z}\times\nabla n_{so}\cdot\nabla\phi + n_{so}\frac{\partial}{\partial_z}v_{sz} = 0$$

For pair-ion plasma, which can also be re-written as

$$d_t\left(Z_+ n_{+1} - Z_- n_{-1}\right) - \frac{c}{B_o}\hat{z}\times\nabla\left(Z_+ n_{+o} - Z_- n_{-o}\right)\cdot\nabla\phi + \frac{\partial}{\partial_z}\left(Z_+ n_{+o}v_{+z} - Z_- n_{-o}v_{-z}\right) = 0 \tag{37}$$

where $d_t = \partial_t + \frac{c}{B_o}\left(\hat{z}\times\nabla\phi\cdot\nabla\right)$. Using the Poisson's equation and Boltzmann distribution for the electrons, the above equation can approximately be written as,

$$\left[\frac{\alpha^2}{u^2}\left\{Z_+\alpha_o c_{s+}^2\left(1 - \left(\eta_+ - \frac{2}{3}\right)\frac{\tilde{V}_{n+}}{u}\right) - Z_-\beta c_{s-}^2\left(1 - \left(\eta_- - \frac{2}{3}\right)\frac{\tilde{V}_{n-}}{u}\right)\right\} + \frac{u_{ne}}{u} - 1\right]\frac{\partial\Phi}{\partial\xi} + \frac{u_{ne}}{u}\Phi\frac{\partial\Phi}{\partial\xi} + \left(1 - \frac{5\alpha^2 T_+}{3u^2 m_+}\right)\lambda_{De}^2\frac{\partial^3\Phi}{\partial\xi^3} = 0 \tag{38}$$

Since in most of laboratory produced plasmas, the mass of $m_+$ and $m_-$ ions is approximately same. i.e., $m_+ \simeq m_-$. Furthermore, the temperature of pair-ion is slightly different i.e., $(0.3 - 0.5)$ eV, one can may take $T_+ \simeq T_-$. Since, we are considering a special case in which $m_- > m_+$; $T_{+o} > T_{-o}$; such that $T_{+o}/m_+ > T_{-o}/m_-$, thus we can write $T_{+o}/m_+ =$ $(1+\epsilon)T_{-o}/m_-$, whereas, in most of the experiments $\epsilon > 0$ but is a small number. To obtain an approximately good result, in the last term of Eq. (38), we take $\epsilon$ very small we finally obtain,

$$a_1 \frac{\partial \Phi}{\partial \xi} + a_2 \Phi \frac{\partial \Phi}{\partial \xi} + a_3 \frac{\partial^3 \Phi}{\partial \xi^3} = 0$$

where

$$a_1 = \left[1 + \frac{u_{ne}}{u} - \frac{\alpha^2}{u^2}\left\{\frac{5}{3\tau_+}c_{s+}^2 + Z_+\alpha_o c_{s+}^2\left(1 - \left(\eta_+ - \frac{2}{3}\right)\frac{\tilde{V}_{n+}}{u}\right) - Z_-\beta c_{s-}^2\left(1 - \left(\eta_- - \frac{2}{3}\right)\frac{\tilde{V}_{n-}}{u}\right)\right\}\right]$$

$$a_2 = \frac{u_{ne}}{u}$$

and

$$a_3 = \left(\frac{5\alpha^2}{3u^2\tau_+}c_{s+}^2 - 1\right)\lambda_{De}^2$$

The above equation is the well known Korteweg-de Vries (KdV) type equation which admits a localized pulse type soliton solution such that

$$\Phi = \Phi_m \sec^2 h\left[\frac{\xi}{\Delta}\right], \tag{39}$$

where $\Phi_m = 3u/a_{21}$, $\Delta = \sqrt{4a_{31}/u}$ with $a_{21} = a_2/a_1$ and $a_{31} = a_3/a_1$. The above solution would be valid for $a_{31} > 0$, which means $\left[1 + \frac{u_{ne}}{u} - \frac{\alpha^2}{u^2}\left\{\frac{5}{3\tau_+}c_{s+}^2 + Z_+\alpha_0 c_{s+}^2\left(1 - \left(\eta_+ - \frac{2}{3}\right)\frac{\tilde{V}_{n+}}{u}\right) - Z_-\beta c_{s-}^2\left(1 - \left(\eta_- - \frac{2}{3}\right)\frac{\tilde{V}_{n-}}{u}\right)\right\}\right] > 0$. In this paper, we have obtained ITG driven solitons for pair-ion plasma with electrons follow Boltzmann type distribution.

## V. CASE 3 (NON MAXWELLIAN PAIR ION ELECTRON PLASMA)

Solving Eqs. (5) - (7) for pair-ion plasma, we obtain the following dispersion relation under the assumption that the perturbation wavelength is much larger than the electron Debye shielding length such that $k\lambda_{De} \ll 1$, and obtain the following relation,

$$\left(\omega - \frac{5}{3}\omega_{B+}\right)\left(\omega - \frac{5}{3}\omega_{B-}\right)\left(Z_+\tau_+\omega^2 - Z_+\tau_+\omega_{B+}\omega - k_z^2c_{s+}^2\right)\left(Z_-\tau_-\omega^2 - Z_-\tau_-\omega_{B-}\omega - k_z^2c_{s-}^2\right)$$

$$= \frac{Z_+\alpha_o}{\beta}\left[\left[\left\{\begin{array}{c} \tau_+^2Z_+^2\left(\omega_{n+}+\omega_{B+}\right)\left(\omega - \frac{5}{3}\omega_{B+}\right)\omega - \tau_+^2Z_+^2\omega_{n+}\omega_{B+}\left(\eta_+ - \frac{2}{3}\right)\omega + \\ k_z^2c_{s+}^2\tau_+\left(\omega - \frac{5}{3}\omega_{B+} - Z_+\omega_{n+}\left(\eta_+ - \frac{2}{3}\right)\right) \\ \left(Z_-\tau_-\omega^2 - Z_-\tau_-\omega_{B-}\omega - k_z^2c_{s-}^2\right)\left(\omega - \frac{5}{3}\omega_{B-}\right) \end{array}\right\}\right]\right]$$

$$- \frac{Z_-\beta_o}{\beta}\left[\left[\left\{\begin{array}{c} \tau_-^2Z_-^2\left(\omega_{n-}+\omega_{B-}\right)\left(\omega - \frac{5}{3}\omega_{B-}\right)\omega - \tau_-^2Z_-^2\omega_{n-}\omega_{B-}\left(\eta_- - \frac{2}{3}\right)\omega + \\ k_z^2c_{s-}^2\tau_-\left(\omega - \frac{5}{3}\omega_{B-} - Z_-\omega_{n-}\left(\eta_- - \frac{2}{3}\right)\right) \\ \left(Z_+\tau_+\omega^2 - Z_+\tau_+\omega_{B+}\omega - k_z^2c_{s+}^2\right)\left(\omega - \frac{5}{3}\omega_{B+}\right) \end{array}\right\}\right]\right] \quad (40)$$

where $Z_+n_{+o}(Z_-n_{-o})$ satisfies the quasineutrality condition of the plasma such that we can write $Z_+n_{+o} - Z_-n_{-o} = n_{eo}$, for pair-ion electron system. Here $\alpha_o = n_{+o}/n_{eo}$ and $\beta_0 = [(1 - Z_+\alpha_o)]$ .Eq. (40) represents a sixth order dispersion relation and its analytical solution is extremely difficult to obtain. We shall therefore discuss numerical results in later section. We shall first present here, some interesting limiting cases in the following sub-section.

## A. Limiting Cases:

### 1. *Case-1:*

First, we consider a uniform density and magnetic field case, for which we can take $\omega_{n+} = \omega_{n-} = 0$; $\omega_{B+} = \omega_{B-} = 0$,and $\eta_+ = \eta_- = 0$. If we further assume that the phase velocity of wave lies in the rage of ion-acoustic and ion thermal speed such that: $v_{ts}^2 < \frac{\omega^2}{k^2} < c_{ss}^2$ , we obtain the following dispersion relation:

$$\omega^2 = k_z^2\left[\frac{\alpha_0}{\beta}c_{s+}^2 + \frac{\beta_o}{\beta}c_{s-}^2\right] \quad (41)$$

where $\alpha_0$ and $\beta_o$ are defined earlier. The above dispersion relation represents a modified coupled ion-acoustic mode with variable mass, charge state and number density of positive and negative ions.

*2. Case-2:*

Secondly, we consider a nonuniform pair-ion plasma having equilibrium density and temperature gradients such that $\omega_{ns}, \neq 0, \eta_s \neq 0$. To investigate coupled ion-temperature-gradient- driven drift mode with ion-acoustic mode for pair-ion plasma, we assume the phase velocity of the wave in the range $c_{ss}^2 > \frac{\omega^2}{k^2} > v_{ts}^2$ and keeping magnetic field uniform for simplification, we obtain

$$\omega^3 = k_z^2 \left[\frac{\alpha_o}{\beta} c_{s+}^2 \omega + \frac{\beta_o}{\beta} c_{s-}^2 \omega\right] + Z_+ \frac{\alpha_o}{\beta}\left(\omega^2 \tau_+ Z_+ \omega_{n+} + c_{s+}^2 k_z^2 \omega_{n+}\left(\eta_+ - \frac{2}{3}\right)\right)$$
$$- Z_- \frac{\beta_o}{\beta}\left(\omega^2 \tau_- Z_- \omega_{n-} + c_{s-}^2 k_z^2 \omega_{n-}\left(\eta_- - \frac{2}{3}\right)\right) \tag{42}$$

which is third order dispersion relation, essentially representing a coupled ion-acoustic type mode with ion-temperature-gradient driven drift mode.

### B. Nonlinear Solution

Solving the system of non Maxwellian pair ion electron plasma we finally obtain

$$a_1 \frac{\partial \Phi}{\partial \xi} + a_2 \Phi \frac{\partial \Phi}{\partial \xi} + a_3 \frac{\partial^3 \Phi}{\partial \xi^3} = 0 \tag{43}$$

where $a_1 = \left[\beta\left(\frac{5\alpha^2 T_+}{3u^2 m_+} - 1\right) + \frac{u_{ne}}{u} - \frac{\alpha^2}{u^2}\left\{\begin{array}{c} Z_+ \alpha_o c_{s+}^2 \left(1 - \left(\eta_+ - \frac{2}{3}\right)\frac{\tilde{V}_{n+}}{u}\right) - \\ Z_- \beta_o c_{s-}^2 \left(1 - \left(\eta_- - \frac{2}{3}\right)\frac{\tilde{V}_{n-}}{u}\right)\end{array}\right\}\right]$, $a_2 = \beta \frac{u_{ne}}{u}$ and $a_3 = \left(\frac{5\alpha^2}{3u^2 \tau_+} c_{s+}^2 - 1\right)\lambda_{De}^2$. The above equation is the well known Korteweg-de Vries (KdV) type equation which admits a localized pulse type soliton solution such that

$$\Phi = \Phi_m \sec^2 h\left[\frac{\xi}{\Delta}\right],$$

where $\Phi_m = 3u/a_{21}$, $\Delta = \sqrt{4a_{31}/u}$ with $a_{21} = a_2/a_1$ and $a_{31} = a_3/a_1$. The above solution would be valid for $a_{31} > 0$, which means $\left[\beta\left(\frac{5\alpha^2 T_+}{3u^2 m_+} - 1\right) + \frac{u_{ne}}{u} - \frac{\alpha^2}{u^2}\left\{\begin{array}{c} Z_+ \alpha_o c_{s+}^2 \left(1 - \left(\eta_+ - \frac{2}{3}\right)\frac{u_{n+}}{u}\right) - \\ Z_- \beta_o c_{s-}^2 \left(1 - \left(\eta_- - \frac{2}{3}\right)\frac{u_{n-}}{u}\right)\end{array}\right\}\right] > 0$. In this paper, we have obtained ITG driven solitons for pair-ion plasma with electrons following Boltzmann type distribution.

## VI. NUMERICAL RESULTS AND DISCUSSION

In this section, we present the numerical results by choosing some typical parameters of fullerene pair ion plasma with non thermal electron distribution. For fullerene plasma, we choose $n_i = 2 \times 10^7 cm^{-3}$, $B_o = 0.3G$, $T_e = 1eV$, $T_+ = 0.3eV$, $T_- = \frac{T_+}{1.5}$ . In Figures (1)-(4) effect of growth rate of ion-temperature-gradient (ITG) instability for pair ion plasma with nonthermal electron distribution has been shown for Eq. (40). It can be seen that in Figure (1), by varying $\eta = 7, 8, 9$ and keeping other parameters fixed, growth rate of instability increases. Figure (2) has been plotted for different temperature ratios, that is for $T_- = \frac{T_+}{1.5}$ and for equal temperature pair-ions $T_+ = T$ . It is evident from the graph that growth rate of instability is slightly higher than that of equal temperature pair-ion case. In Figure $(3)-(4)$, imaginary part of frequency is plotted versus $k_z$ for different values of kappa ($\kappa = 3, 4, 10$) and ($\Gamma = 0.20,\ 0.30,\ 0.40$). It can be seen that instability increases as nonthermal electron population increases.

In Figure (5) and (6), we have plotted Eq. (39) which clearly shows solitary wave structure ITG driven case for different parameters of pair-ion plasma system. For instance, in Figure (5) shows solitary wave potential $\Phi$ varies as a function of $\xi$ for different ratios of ion to electron number densities such that we take $\alpha_o = 0.75,\ 0.85,\ 0.95$ for fixed values of $\eta_+ = \eta_- = \eta = 9$. It is evident from the graph that the amplitude of soliton increases with the increase of parameter $\alpha_o$. Figure (6)-(7) depicts the variation of soliton amplitude by varying $\kappa$ and $\Gamma$. It shows that, by increasing nonthermal electron population ($\kappa = 3, 4, 10$) and ($\Gamma = 0.20,\ 0.30,\ 0.40$), amplitude of solitary wave increases.

## VII. SUMMARY

In this paper, we have investigated linear and nonlinear dynamics of ion-temperature-gradient driven drift mode for maxwellian and non maxwellian pair-ion plasma, maxwellian and non maxwellian pair-ion electron plasma embedded in an inhomogeneous magnetic field having gradients in ion's temperature and number density. Linear dispersion relations are derived and analyzed analytically as well as numerically for all these cases. It has been found that growth rate of instability increases with increasing $\eta$.In the nonlinear regime, soliton structures are found to exist. Our numerical analysis shows that amplitude of solitary waves increases by increasing ion to electron number density ratio. These solitary structures are also found to be sensitive to non thermal kappa and Cairns distributed electrons. Our present work may contribute a good illustration of the observation of nonlinear solitary waves driven by the ITG mode in magnetically confined pair-ion plasmas and space pair-ion plasmas as the formation of localized structures along drift modes is one of the striking reasons for L-H transition in the region of improved confinements in magnetically confined devices like tokamaks.

**Achnowledgemnts**

One of us (Z.E.) is grateful to the office of CAAD, National Center for Physics (NCP) -Islamabad and theoretical physics group at QAU for the hospitality where this work was initiated. J.R. is thankful to Zahida Ehsan for her visits to SPAR-CUI Lahore where this work was completed.

**Authors' contribution**

The idea was floated by Z.E, J.R and A.M.M made analytical derivations. All authors contributed equally to the results and discussions sections. Whereas write-up was done by Z.E.

---

# Figure Captions

Fig. 1: (Color Online) Plot of imaginary part of frequency versus $k_z$ for different values of $\eta_+ = \eta_- = \eta = 3, 5, 9$, keeping other parameters fixed.

Fig. 2: (Color Online) Plot of imaginary part of frequency versus $k_z$ for $T_- = \frac{T_+}{1.5}$ and $T_+ = T_-$, keeping other parameters fixed.

Fig. 3: (Color Online) Plot of imaginary part of frequency versus $k_z$ with different values of kappa ($\kappa = 3, 4, 10$).

Fig. 4: (Color Online) Plot of imaginary part of frequency versus $k_z$ with different values of gamma ($\Gamma = 0.20,\ 0.30,\ 0.40$).

Fig. 5: (Color Online)Effect of ratio of ion to electron number densities ($\alpha_o = 0.75,\ 0.85,\ 0.95$) on soliton.

Fig. 6: (Color Online) Effect of nonthermality parameter $\kappa(\kappa = 3,\ 4,\ 10)$ on soliton profile.

Fig. 7: (Color Online) Effect of nonthermality parameter $\Gamma(\Gamma = 0.20,\ 0.30,\ 0.40)$ on soliton profile.

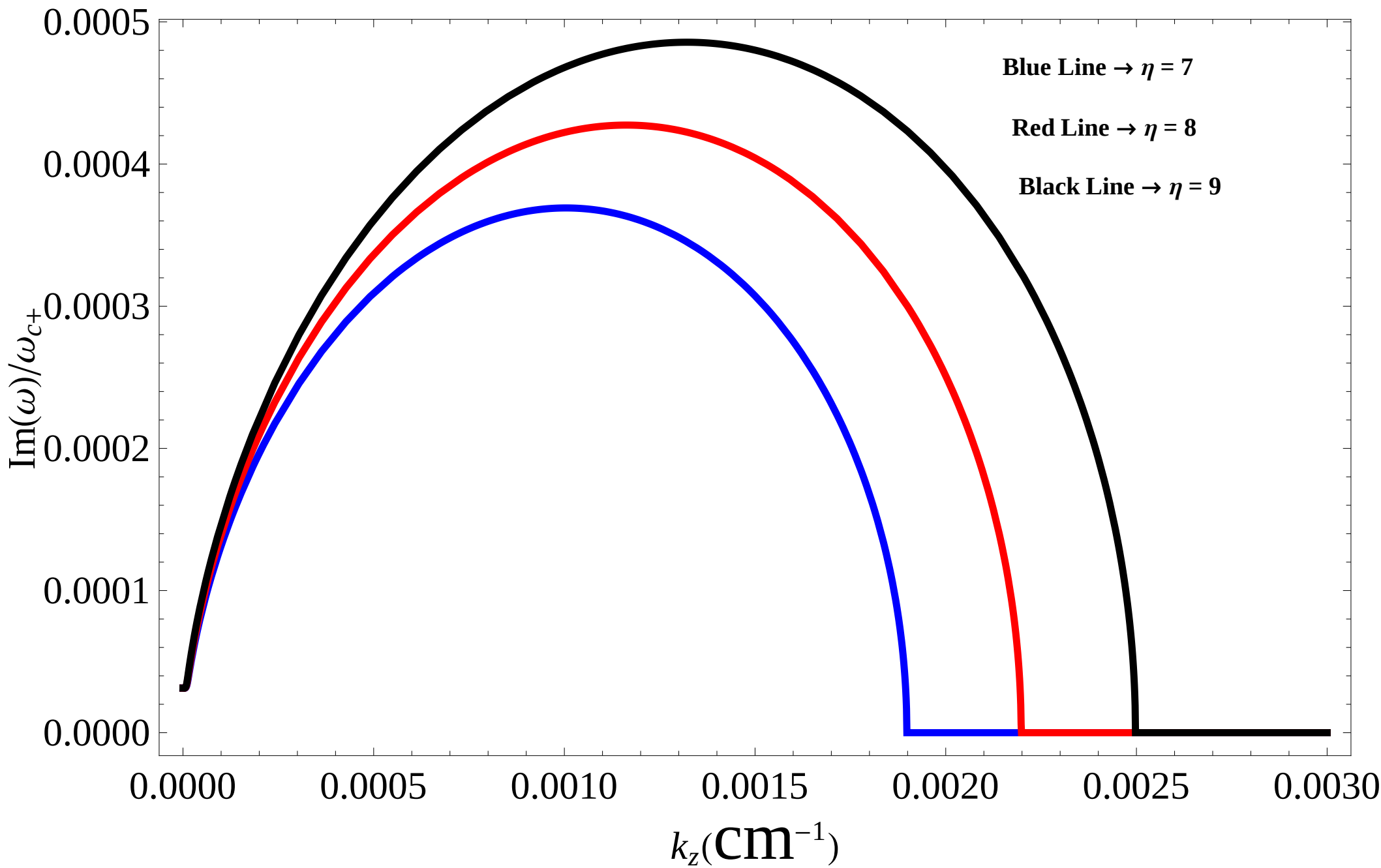

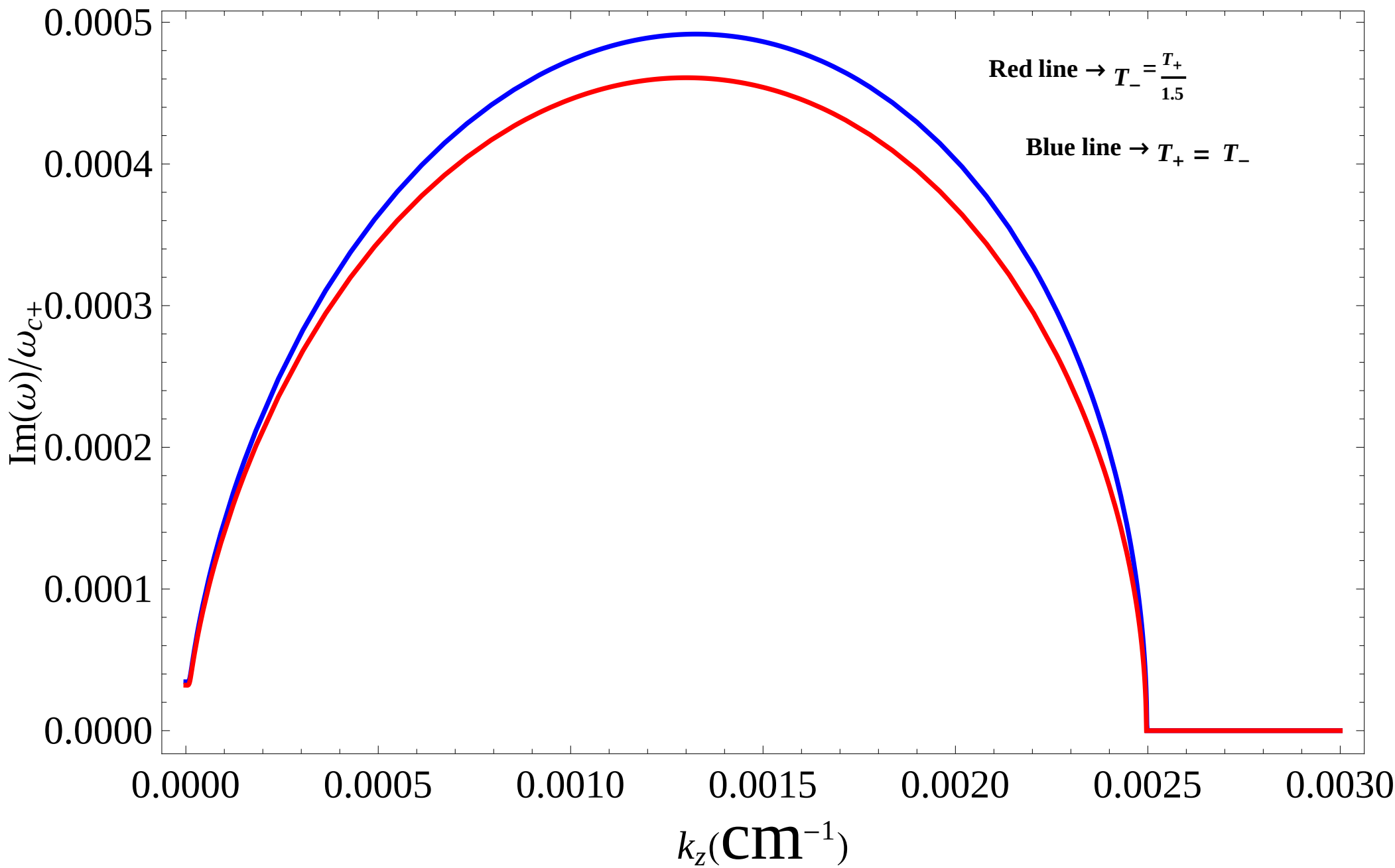

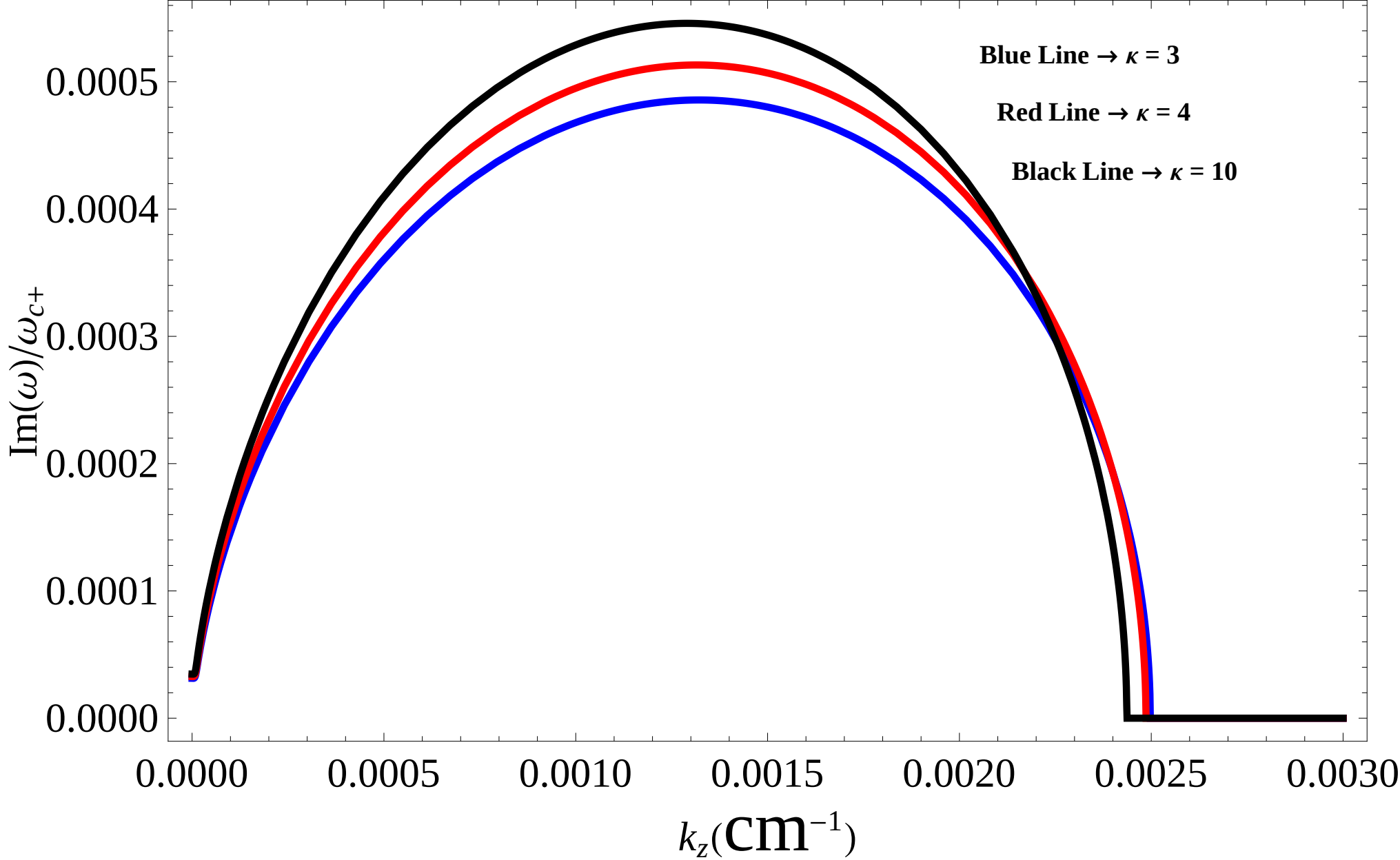

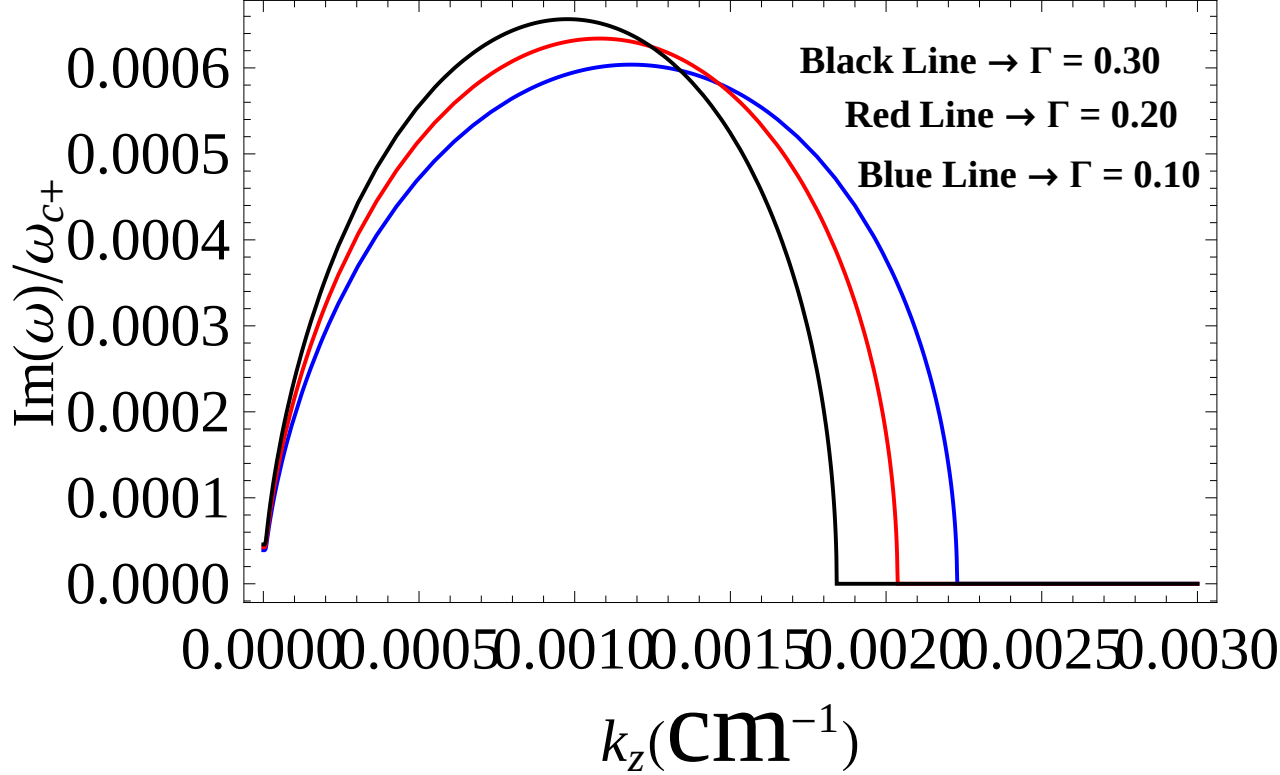

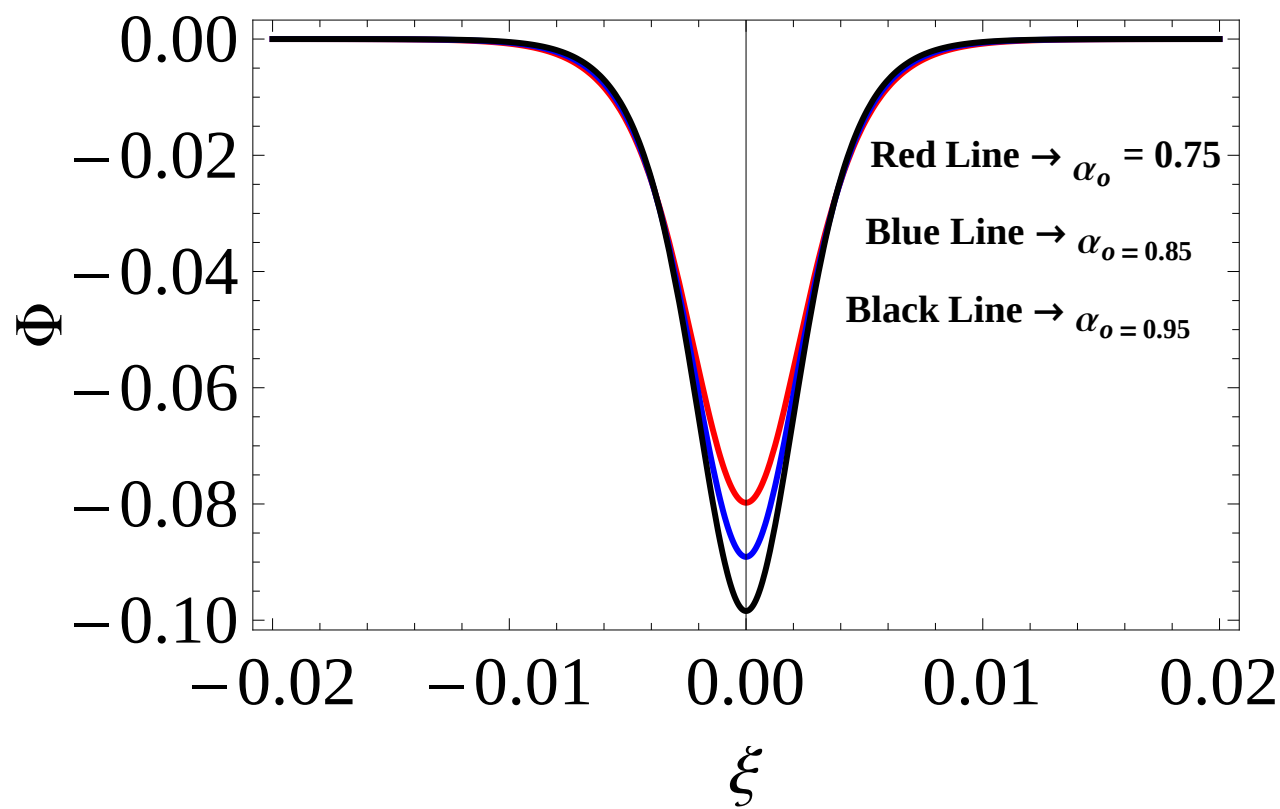

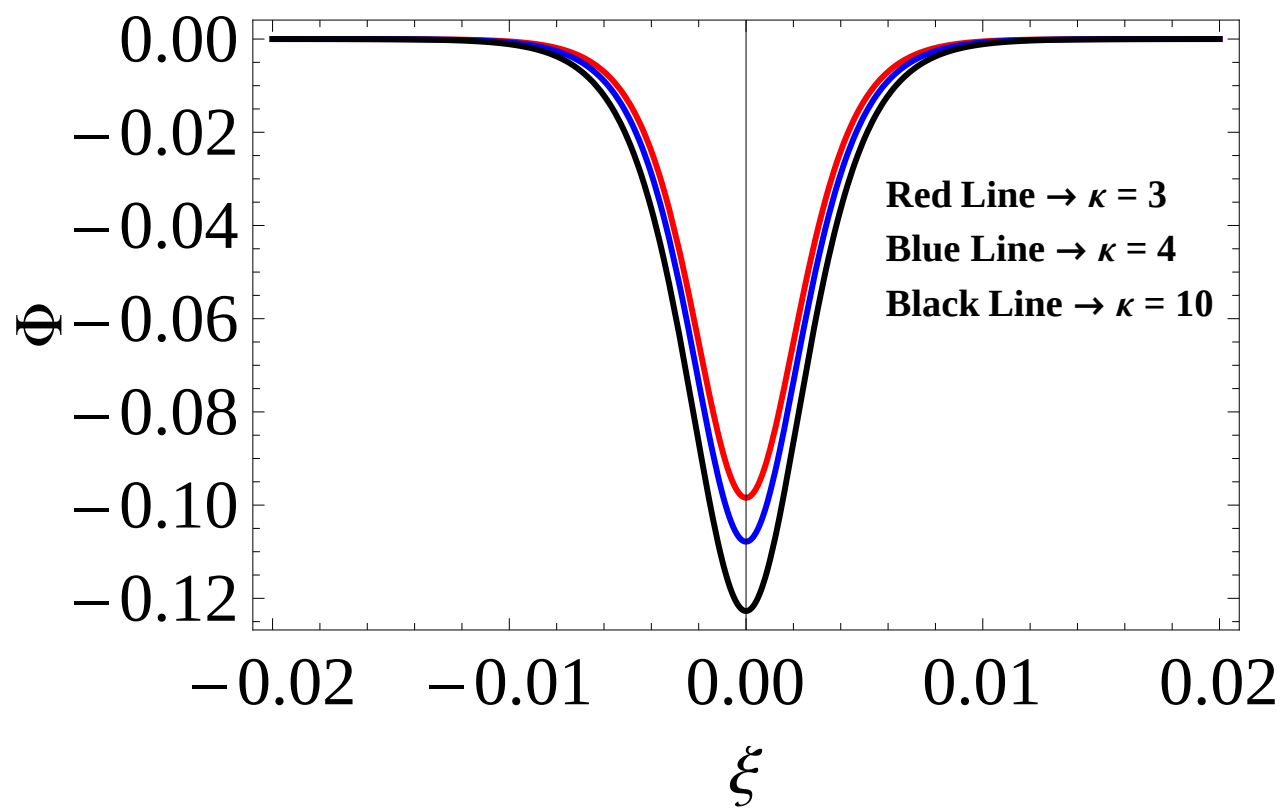

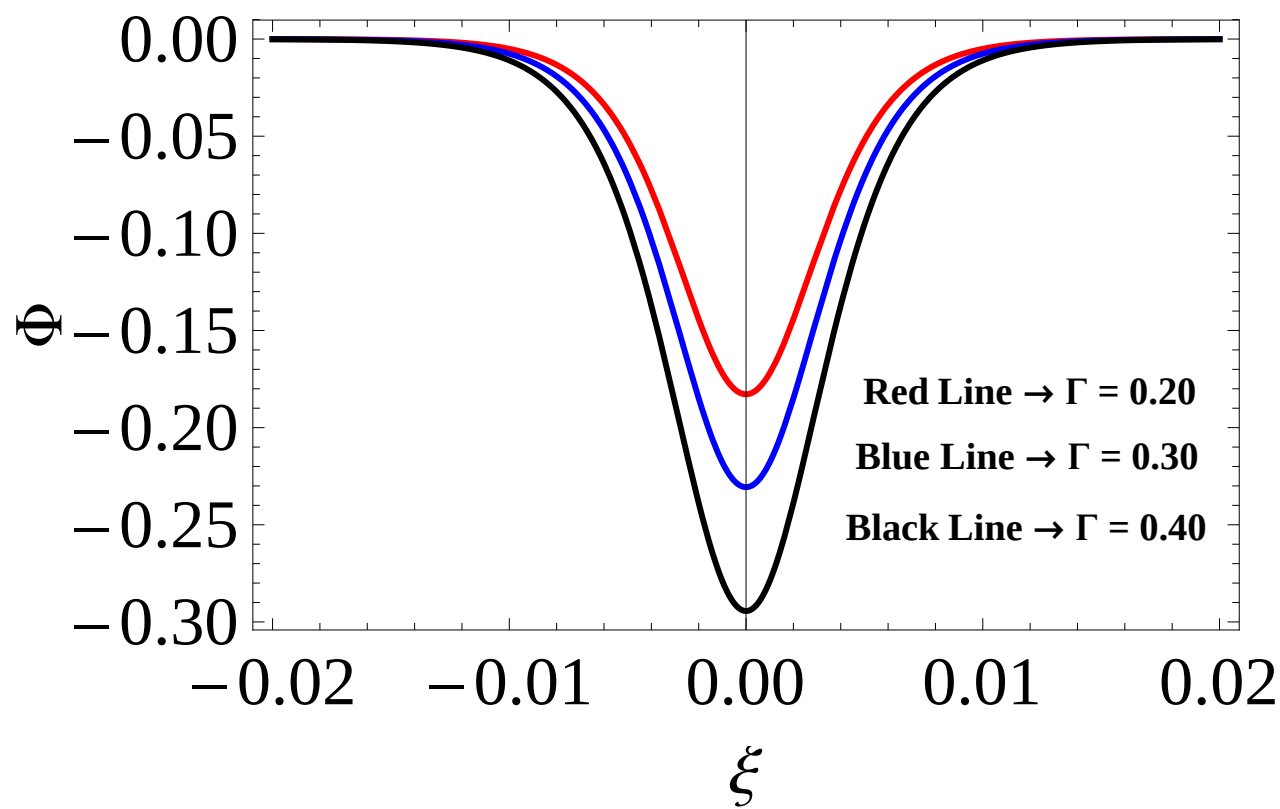